\documentstyle[11pt,newpasp_mod,twoside,epsf]{article}
\newcommand{\feh}{\mbox{[Fe/H]}}
\newcommand{\Msun}{\mbox{$M_{\odot}$}}

\newcommand{\sub}[1]{\mbox{$_{\rm #1}$}}

\newcommand{\Teff}{\mbox{$T\sub{eff}$}}

\newcommand{\comment}[1]{}

\def\edcomment#1{\iffalse\marginpar{\raggedright\sl#1\/}\else\relax\fi}
\reversemarginpar

\begin{document}
\title{Theoretical expectations for clump red giants as distance indicators}
 \author{L\'eo Girardi}
\affil{Dipartimento di Astronomia, Vicolo dell'Osservatorio 5, I-35122
Padova, Italy}

\begin{abstract}
Variations of $\sim0.4$~mag are expected in the $I$-band absolute 
magnitude of red clump giants, $M_I^{\rm RC}$, 
as a function of both stellar age and metallicity. This
regardless of some potential theoretical uncertainties. 
Due to the quite large differences in mean ages and 
metallicities of clump stars among galaxies, systematic changes 
(also amounting up to $\sim0.4$~mag) come out in their $M_I^{\rm RC}$. 
These numbers also indicate a distance to the LMC that 
is not necessarily ``short''.
\end{abstract}

\subsection*{Introduction}
Any measurement of distance modulus using red clump stars 
requires the determination of four quantities:
	\begin{equation}
(m-M)_0 = I^{\rm RC}({\rm galaxy}) - M_I^{\rm RC}({\rm Hipp}) - 
	A_I + \Delta M_I^{\rm RC}.
	\end{equation}
The first two -- the apparent clump magnitude in an external galaxy,
$I^{\rm RC}({\rm galaxy})$, and its absolute magnitude in the Solar 
Neighbourhood sampled by {\em Hipparcos}, 
$M_I^{\rm RC}({\rm Hipp})$ -- can be accurately measured,
with $1\sigma$ errors typically $<0.03$~mag (e.g.\
Paczy\'nski \&  Stanek 1998). Differently, 
determinations of the total absorption, $A_I$, are often 
controversial at the level of about
0.2~mag, as demonstrated by the case of the LMC field population
(Udalski 1998; Romaniello et al. 1999; Zaritsky 1999).
Even more controversial is the assessment of 
the intrinsic differences in magnitude
between {\em Hipparcos} and the external 
populations of clump stars, $\Delta M_I^{\rm RC}$.
In this regard, the dependence 
of $M_I^{\rm RC}$ on age and metallicity has  been claimed
to be either (i) small and empirically-calibrated 
(Udalski 1998ab, 2000), or (ii) more
significant as suggested by stellar models (Cole 1998;
Girardi et al.\ 1998) and observations of open clusters 
(Twarog et al.\ 1998; Sarajedini 1998). 
Moreover, Girardi \& Salaris (2000) conclude
that present empirical calibrations for 
$\Delta M_I^{\rm RC}$ (Udalski 1998ab, 2000) 
do not represent general relations, and 
hence are not suitable for being used in eq.~(1). 
The discussion about $A_I$ and $\Delta M_I^{\rm RC}$
is especially important in the case of the LMC, because 
the red clump method has been claimed to provide strong evidence
for a ``short'' distance scale (Udalski 2000, and references 
therein).

In the limited space of this review, I just recall the main reasons 
why theoretical models predict non-negligible (up to 0.4~mag) 
$\Delta M_I^{\rm RC}$ values. For a detailed discussion -- 
based on the same kind of analysis as here -- of the 
related observational data, I refer to Girardi \&  Salaris (2000).

\subsection*{Why $M_I^{\rm RC}$ changes with age and metallicity}
Firstly, I recall that in the interval of effective
temperatures that characterize red clump stars ($3.8\la\Teff\la3.6$;
see Girardi et al.\ 1998), bolometric corrections in the $I$-band
are almost constant (to within 0.1~mag) and depend very little on 
metallicity. This means that $M_I^{\rm RC}$
reflects very well the behaviour of $\log L$, a quantity 
directly predicted by stellar evolution models.

And stellar models have since long predicted that core He-burning
stars (CHeB) of low-mass (say $M<2$~\Msun),
for a given metallicity, should cover a small 
interval of $\log L$ -- as pointed out by e.g.\ Cannon (1970) and 
Castellani et al.\ (1992) -- as a consequence of the similar 
core masses $M_{\rm core}$ that they have at the 
He-flash. In fact, similar $M_{\rm core}$ values imply 
nearly the same luminosities coming from their
He-burning cores, i.e.\ $L_{\rm He} \approx {\rm const}$. 
However, it has not been much emphasized that, 
for the same $M_{\rm core}$ and metallicity, 
the luminosity of the H-burning shell
increases monotonicaly with the envelope mass, so that
$L_{\rm H} \propto M$ (roughly). 
Since $L=L_{\rm H}+L_{\rm He}$ with 
$L_{\rm H}\approx 2\,L_{\rm He}$ at $M=1$~\Msun, it follows that 
low-mass clump stars {\em must} become some tenths of a magnitude
brighter at larger masses/smaller ages. Another trend is that
$L$ should decrease with \feh, due to the decrease of 
both $M_{\rm core}$ and efficiency of the H-burning shell.

Some CHeB intermediate-mass stars do also occupy the 
clump region of the HR diagram (Girardi 1999). 
Their luminosities change a lot with age, reflecting mainly the 
proportionality between the core mass at He-ignition and the 
initial mass, that is followed by stars that do not develop 
e$^{-}$-degenerate He cores.
\begin{figure}
\plotone{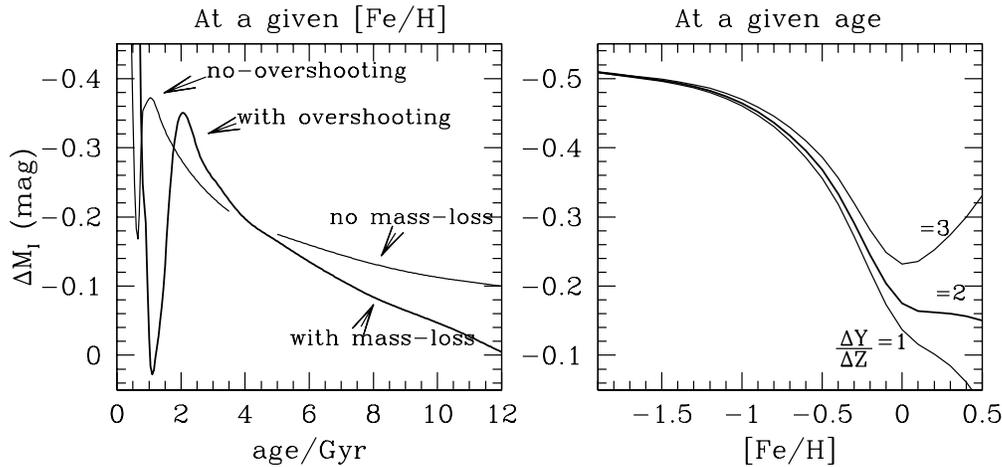}
\caption{$M_I^{\rm RC}$ as a function of age and \feh\ (schematic).}
\end{figure}

These behaviours are shared by most (if not all) sets of 
evolutionary tracks of CHeB stars present in the
literature (e.g.\ Pols et al.\ 1998; Charbonnel et al.\ 1996; 
Girardi et al.\ 2000, among others), and are schematically 
presented in  Fig.~1: $M_I^{\rm RC}$ is expected to 
change by as much as $\sim0.4$~mag with age at a fixed 
\feh, and by $\sim0.4$~mag with \feh\ at a fixed age. 
A few theoretical uncertainties (e.g.\ overshooting, mass-loss, 
the helium-to-metal enrichment ratio, as shown in Fig.~1) 
can change the details of this behaviour by $\sim0.1$~mag, 
but cannot change the general trends.

\subsection*{Why $M_I^{\rm RC}$ changes among galaxies}
Assessing the changes of $M_I^{\rm RC}$ with age and metallicity is
only half of the problem in determining the $\Delta M_I^{\rm RC}$ 
values to be used in eq.~(1). 
The other half corresponds to answering ``What are 
the age and metallicity distributions of clump stars in 
galaxies\,?''
\begin{figure}
\plotone{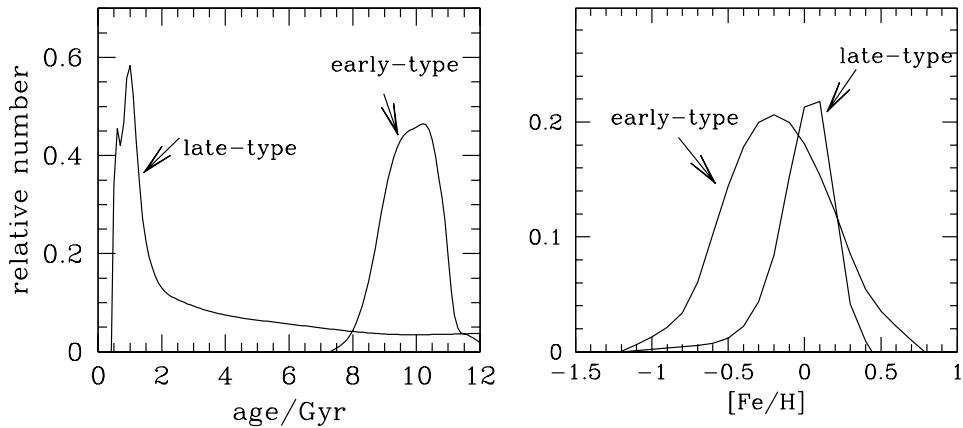}
\caption{Age and \feh\ distributions for clump stars in
late- and early-type galaxies (schematic).}
\label{fig2}
\end{figure}
This question finds quantitative answers from basic population 
synthesis theory, once we know the past star formation rate (SFR) and 
age-metallicity relation (AMR) of the galaxies considered 
(see Girardi \& Salaris 2000). Fig.~2 shows two illustrative
cases:

(i) In {\bf a late-type galaxy} that has
formed stars at a nearly constant rate since its
formation until now (e.g.\ the disks of spirals, or irregulars), 
the age distribution of clump stars strongly
concentrates at relatively young ages ($1-3$~Gyr). 
This effect comes out because: (i) the rate
at which stars leave the main sequence decreases with age; and (ii)
the He-burning lifetime has a maximum for stars of
$\sim1$~Gyr. Notice that a predominantly young clump implies a
somewhat narrow \feh\ distribution, since, in general,
galaxies have experienced little chemical evolution in
the last few gigayears.

(ii) In {\bf an early-type galaxy} where star formation ceased 
after just a few Gyr (e.g.\ ellipticals and the bulges of spirals),
clump stars younger than $\sim8$~Gyr simply do not exist
and the clump age distribution is narrow. These old populations, 
in general, present a broader distribution of \feh\ values. 

Of course, any intermediate (and even more extreme) behaviour can
occur. The general situation is
that {\em quite different distributions of ages/metallicities of clump
stars are expected among different galaxies.}
The local sample of stars that define the Hipparcos clump, is also
expected to be quite different from those found in other Local Group 
systems.

\subsection*{Conclusions about $\Delta M_I^{\rm RC}$}
If we consider Figs.~1 and 2 together,
the conclusions are almost immediate: {\em non-neglegible
values of $\Delta M_I^{\rm RC}$ are expected among different galaxies}.
Girardi et al.\ (1998) and Girardi \& Salaris (2000) find 
$\Delta M_I^{\rm RC}$ values as large as $0.2-0.3$~mag for the 
Magellanic Clouds. When one uses these theoretical
population corrections, and the Romaniello et al.\ (1999) or
Zaritsky (2000) determinations for $A_I$, it turns out that the 
red clump method is still compatible with a ``long'' 
distance scale, with $(m-M)_0^{\rm LMC}=18.55\pm0.05$~mag 
for the LMC.

These conclusions seem to be in contradiction with  
Udalski (1998ab, 2000), who, on observational grounds, finds
a small dependence of $M_I^{\rm RC}$ on both age and metallicity.
However, Girardi \& Salaris (2000) clarify that (i)
larger dependences, compatible with theoretical models, 
are found by Twarog et al.\ (1998) and Sarajedini (1999);
(ii) Udalski's empirical relations do not have general validity,
since they reflect the particular distributions of ages and 
metallicities of clump stars included in the 
observational samples, and express the clump 
behaviour by means of a too simple relation between 
$M_I^{\rm RC}$ and \feh\ 
(which, cf.\  Figs.~1 and 2, is not expected to be 
satisfactory, for both star clusters and galaxies). 
According to Girardi \& Salaris, theoretical models still 
provide the most reliable $\Delta M_I^{\rm RC}$ values, 
provided that the SFR and AMR of the galaxies under scrutiny 
are sufficiently well known. 

In summary, theoretical predictions seem to 
survive to most tests provided by observational data 
(cf.\ Girardi \& Salaris 2000), and yet lead to an 
interpretation of the data that is substantially different -- 
and much more complete\,! -- than the one sketched in present 
empirical $M_I^{\rm RC}$ versus \feh\ calibrations. 
Can we really do without theoretical models\,?

\paragraph{Acknowledgements}
I am greatly indebted to Maurizio Salaris, Martin Groenewegen 
and Achim Weiss,
my collaborators in the clump work. 


\end{document}